\documentclass[12pt,preprint]{aastex}

\def\max{{\rm max}}

\def\eff{{\rm eff}}

\begin{document}
\title{The KMTNet 2016 Data Release}

\author{\textsc{
H.-W. Kim$^{1}$, 
K.-H. Hwang$^{1}$,  
D.-J. Kim$^{1}$, 
M. D. Albrow$^{2}$,
S.-M. Cha$^{1,3}$, 
S.-J. Chung$^{1,4}$, 
A. Gould$^{1,5,6}$, 
C. Han$^{7}$, 
Y. K. Jung$^{1}$, 
S.-L. Kim$^{1,4}$, 
C.-U. Lee$^{1,4}$,
D.-J. Lee$^{1}$,
Y. Lee$^{1,3}$, 
B.-G. Park$^{1,4}$,
R. W. Pogge$^{5}$ 
Y.-H. Ryu$^{1}$, 
I.-G. Shin$^{8}$, 
Y.~Shvartzvald$^{9,^{\dag}}$, 
J. C. Yee$^{8}$, 
W.~Zang$^{10,11}$, 
W. Zhu$^{12}$, 
\\
(KMTNet Collaboration)\\}}

\affil{$^{1}$Korea Astronomy and Space Science Institute, Daejon
34055, Republic of Korea}

\affil{$^{2}$University of Canterbury, Department of Physics and
Astronomy, Private Bag 4800, Christchurch 8020, New Zealand}

\affil{$^{3}$School of Space Research, Kyung Hee University,
Yongin, Kyeonggi 17104, Republic of Korea}

\affil{$^{4}$Korea University of Science and Technology, 
217 Gajeong-ro, Yuseong-gu, Daejeon 34113, Republic of Korea}

\affil{$^{5}$Department of Astronomy, Ohio State University, 140 W.
18th Ave., Columbus, OH 43210, USA}

\affil{$^{6}$Max-Planck-Institute for Astronomy, K\"{o}nigstuhl 17,
69117 Heidelberg, Germany}

\affil{$^{7}$Department of Physics, Chungbuk National University,
Cheongju 28644, Republic of Korea}

\affil{$^{8}$Harvard-Smithsonian Center for Astrophysics, 60 Garden St.,Cambridge, MA 02138, USA}

\affil{$^{9}$Jet Propulsion Laboratory, California Institute of
Technology, 4800 Oak Grove Drive, Pasadena, CA 91109, USA}

\affil{$^{10}$Physics Department and Tsinghua Centre for
Astrophysics, Tsinghua University, Beijing 100084, China}

\affil{$^{11}$Department of Physics, Zhejiang University, Hangzhou,
310058, China}

\affil{$^{12}$Canadian Institute for Theoretical Astrophysics, 
University of Toronto, 60 St George Street, Toronto, ON M5S 3H8, Canada}

\affil{$^{\dag}$NASA Postdoctoral Program Fellow}


\begin{abstract}
We present Korea Microlensing Telescope Network (KMTNet)
light curves for microlensing-event candidates for the 2016 season,
which covers an area of $97\,\deg^2$ observed at cadences
ranging from $\Gamma=0.2\,{\rm hr}^{-1}$ to $\Gamma=8\,{\rm hr}^{-1}$
from three southern sites in Chile, South Africa, and Australia.
These 2163 light curves are comprised of
1856 ``clear microlensing'' and 307 ``possible microlensing'' events
(including 265 previously released from the K2 C9 field).
The data policy is very similar to the one governing the 2015
release.  The changes relative to 2015 in the algorithms to find
and vet microlensing events are comprehensively described.

\end{abstract}

\keywords{gravitational lensing: micro}

\section{{Introduction}
\label{sec:intro}}

A major goal of the Korea Microlensing Telescope Network 
(KMTNet, \citealt{kmtnet}) is to provide timely public access to all 
of the microlensing light curves found by the KMTNet team, including
both ``clear'' and ``possible'' events \citep{eventfinder}.
These events are found by applying the ``event finder'' algorithm
described by \citet{eventfinder} to roughly $3\times 10^8$ light curves
that are derived from data taken by three identical $4\,\deg^2$
cameras mounted on three identical 1.6m telescopes located in 
Chile (KMTC), South Africa (KMTS), and Australia (KMTA).

The first such data release was for the 2015 commissioning-year data set
\citep{eventfinder}.  While the basic goals and data release policy
are very similar to that work, there are many concrete differences
reflected in the 2016 release.  The data set itself is substantially
different (Section~\ref{sec:data}), the event-finder algorithm has
been upgraded, including additional checks for false positives
(Section~\ref{sec:eventfinder}), and the data products have
been substantially upgraded as well (Section~\ref{sec:product}).
The data policy is also slightly altered for 2016 in that a subset
of the data \citep{2016k2} was released in an expedited fashion.
We briefly review these data policies (as slightly modified) in
Section~\ref{sec:policy}.

\section{{2016 Data}
\label{sec:data}}

The 2016 KMTNet data differ in two major respects compared to 2015.
First, we observed a total of 27 fields covering 
$97\,\deg^2$ in 2016 compared to four fields covering $16\,\deg^2$ in 2015.
Second, in 2016, the data from all three sites were reduced and
jointly fit to microlensing light-curve profiles to find candidate
events.  By contrast, in 2015, only KMTC data were initially fit to
microlensing profiles, and the data from KMTS and KMTA were only
reduced for candidates that were identified based on KMTC data.

The layout and nominal  cadence of the 27 fields observed in 2016 is
shown in Figure~12 of \citet{eventfinder}.  Conceptually, there
are three ``prime fields'' covering $12\,\deg^2$ that are observed at 
a cadence of
$\Gamma=4\,{\rm hr}^{-1}$.  In practice, each of these is observed in
pairs (BLG01/BLG41, BLG02/BLG42, BLG03/BLG43) that are offset by about 
$8.4^\prime$ (in order to cover the gaps between the four chips in each
field), all with a nominal per field cadence of $\Gamma=2\,{\rm hr}^{-1}$.  
Moreover, two of these pairs of fields (BLG02/BLG42 and BLG03/BLG43)
overlap by about $0.4\,\deg^2$.  Finally, about $0.35\,\deg^2$ of the
``offset fields'' BLG41 and BLG43 that do not overlap other prime
fields, do overlap the $\Gamma=1\,{\rm hr}^{-1}$ fields (see below) 
BLG04 and BLG22.  Hence the ``prime fields'' actually comprise a total
of about $13.1\,\deg^2$ consisting of 
$0.4\,\deg^2$ with $\Gamma=8\,{\rm hr}^{-1}$,
$9.7\,\deg^2$ with $\Gamma=4\,{\rm hr}^{-1}$,
$0.35\,\deg^2$ with $\Gamma=3\,{\rm hr}^{-1}$, and
$2.65\,\deg^2$ with $\Gamma=2\,{\rm hr}^{-1}$.

The remaining 21 fields covering $84\,\deg^2$ (less the $0.35\,\deg^2$
just mentioned) are divided into three groups: 
7 fields with cadence $\Gamma=1\,{\rm hr}^{-1}$
(BLG04, BLG14, BLG15, BLG17, BLG18, BLG19, and BLG22),
11 fields with cadence $\Gamma=0.4\,{\rm hr}^{-1}$
(BLG11, BLG16, BLG20, BLG21, BLG31, BLG32, BLG33, BLG34,
BLG35, and BLG38), and
3 fields with cadence $\Gamma=0.2\,{\rm hr}^{-1}$
(BLG12, BLG13, and BLG36).  Figure~\ref{fig:starcat} shows the
density of catalog stars in these fields (but excluding BLG41, BLG42, and
BLG43, to avoid clutter).

This nominal cadence was followed by KMTC for all of 2016, and it
was followed for most of 2016 by KMTS and KMTA.  However, this cadence
structure was adjusted to support the {\it Kepler K2}C9 
campaign \citep{gouldhorne,henderson16,2016k2} from April 23
to June 16.  During this period (for KMTS and KMTA), the cadences of fields
BLG02, BLG42, BLG03, and BLG43 were increased by a factor 1.5
from $\Gamma=2\,{\rm hr}^{-1}$ to $\Gamma=3\,{\rm hr}^{-1}$,
while those of all other fields were decreased by a factor 0.75.
That is, the cadences were
$\Gamma=1.5\,{\rm hr}^{-1}$ (2 fields),
$\Gamma=0.75\,{\rm hr}^{-1}$ (7 fields),
$\Gamma=0.3\,{\rm hr}^{-1}$ (11 fields), and
$\Gamma=0.15\,{\rm hr}^{-1}$ (3 fields).

Finally, we note that in 2016, 10/11 of observations
at KMTC, 20/21 of observations at KMTS, and all observations
at KMTA were taken $I$ band.  The remaining observations 
(roughly 5\% of the total) were taken in $V$ band.

\section{{2016 Event Finder}
\label{sec:eventfinder}}

The major modifications to the Event Finder algorithm (originally described
in \citealt{eventfinder}) for 2016 are detailed below. The major change to the
algorithm itself was to combine light curves from multiple sites and
multiple fields (Section~\ref{sec:combined-lc}). 
Another major change was to remove known variables either from 
published catalogs or previous years' work (Section~\ref{sec:known-var}).
We also augmented our automated candidate vetting by writing two new
algorithms whose purposes were to reject ``short-timescale artifacts''
(Section~\ref{sec:short-art}) and ``periodic variables'' 
(Section~\ref{sec:periodic-var}), respectively.  Because
the Event Finder algorithm was continually developed, these algorithms were
only applied to the data analyzed after their development, i.e., they were
incorporated starting approximately at the point just after the prime fields
were completed. Section~\ref{sec:final-vet} describes the final vetting 
procedures and summarizes the discoveries.

\subsection{{Combined Light Curves}
\label{sec:combined-lc}}

The event finder was changed in 2016 in a number of important
respects.  The most important change is that we searched
for events in all three data sets simultaneously.
As discussed by \citet{eventfinder}, light curves were
constructed at the locations of input-catalog stars using
the difference imaging analysis (DIA) code of \citet{wozniak2000}.
Because the input catalog is identical for all three 
observatories, it is straightforward to cross-identify
the light curves.  (This was not possible in 2015 because
there was insufficient computing power to reduce data from
all three sites in a timely way.  Moreover, the commissioning
character of the data would have made combined analysis
quite difficult in any case.)

A second major change was that we combined the analysis of light curves
from overlapping fields.  For example, in the overlap
regions between BLG02/BLG42/BLG03/BLG43, 
a total of 12 light curves were jointly analyzed,
i.e., (3 observatories)$\times$(4 fields).  This was unnecessary in
2015 because there were no overlapping fields.

However, we only combined overlapping fields for catalog stars derived
from the OGLE-III catalog \citep{oiiicat}.  Only these catalog entries
could be unambiguously cross-identified between overlapping fields.
For areas that are not covered by OGLE-III, the star catalog was derived
using DoPhot \citep{dophot} on each field separately.  
Although fields BLG01, BLG02, and BLG03 were moved slightly between 2015 
and 2016, one can gain a good impression of the location of these 
non-OGLE-III areas from Figure~8 of \citet{eventfinder}.  See also
Figure~\ref{fig:starcat} of this paper.

\subsection{{Removing Known Variables and Artifacts}
\label{sec:known-var}}

Another important change was the removal of published variables prior to
the stage of showing the light curves of candidates to the operator.
Recall from \citet{eventfinder} that after individual candidates
are cataloged, they are grouped using a friends-of-friends algorithm,
so that only the ``best'' (in terms of $\Delta\chi^2$) is shown to the
operator.  However, if this ``group leader'' is in a published catalog
of variables, then the entire group is skipped.  For this purpose, we
used the OGLE-long-period-variable \citep{oglelpv} and OGLE-dwarf-nova
\citep{ogledn} catalogs.

Similarly, if the ``group leader'' is identified as an earlier-year
candidate that was flagged as a variable or as an artifact, it is also
not shown.  Of course, for 2016 ``earlier year'' can only refer to
2015 and so only to stars in BLG01, BLG02, BLG03, and BLG04 from that
year.  Moreover, because these fields all moved slightly, we could
only apply this procedure to OGLE-III catalog stars, but in any case
these are the overwhelming majority in these fields.  In fact, in 2016,
we only matched the 2016 ``group leaders'' to the list of 2015 ``group 
leaders'' of variables and artifacts.  A more robust
approach would be to match to all the 2015 stars that were in these
variable and artifact groups, not just the ``leaders''.  Unfortunately
we only realized this when we had completed the review of the
fields that overlapped the 2015 fields.  However, the more robust
approach will be implemented in 2017 and future years.

Finally, we also matched to all previous-year OGLE and MOA events,
as listed on their web pages\footnote{
http://ogle.astrouw.edu.pl/ogle4/ews/ews.html and
http://www.massey.ac.nz/$\sim$iabond/moa/alert2016/alert.php}.
For 2016, these identifications were made {\it after} the event
selection, i.e., at the same time as the selected candidates were
cross-identified with current year (2016) OGLE and MOA events.
However, for 2017 and future years, earlier-year OGLE and MOA events will 
be flagged to the operator's attention at the time that
the candidate is reviewed.
Almost all such matches in 2016 were to OGLE, and the great majority 
of these were
to candidates from more than two years previously.  While we did consider
the possibility that these were repeating microlensing events,
in all cases we concluded that they were cataclysmic variables (CVs)
or other outbursting variables, which happened to look something
like microlensing events in their earliest appearance in OGLE data
and also in the 2016 KMT data.  A number of matches were long timescale
events for which OGLE issued an alert in 2015\footnote{Note that to
avoid prejudicing our selection procedure, we do not match to current-year
(2016) OGLE and MOA events.  By matching to previous-year events during
the selection process, we will introduce
prejudice, but to the very small subclass of events of long-timescale
events that bridge two seasons.}.

\subsection{{Short-Timescale Artifacts}
\label{sec:short-art}}

The great majority of short-timescale artifacts consist of 1--2 isolated
spurious points, usually in a single night, that are then almost perfectly
fit to a microlensing profile (which has two continuous free parameters).
These can occasionally be ``coordinated'' between observatories, in particular
if they are caused by the Moon passing through the bulge.  Because the
fit is ``perfect'', the $\chi^2$ renormalization process described 
by \citet{eventfinder} automatically generates a very high $\Delta\chi^2$.
We search for these artifacts by finding the two data points from each
observatory/field combination that contribute the most to $\Delta\chi^2$ and 
then removing their contribution.  If the event fails the $\Delta\chi^2$
criterion\footnote{$\Delta\chi^2>500$ for prime-field stars that lack
OGLE-III counterparts and $\Delta\chi^2>1000$ for all others (see
\citealt{2016k2}).} after this removal, it is eliminated.
This algorithm was applied to the grouped-candidate list 
for the outlying fields in 2016 but not for the prime fields,
BLG01/02/03/41/42/43, because it was developed after visual inspection
of these fields.  However, for 2017 and future years, it 
will be included directly in the event-finder pipeline.

Prior to implementing the ``short-timescale artifact'' finder described
above, we employed a somewhat less precise method of removing
such short-timescale artifacts.  For each field, and for each of the
four chips, we counted the number of candidates with a given $(t_0,t_\eff)$.
If one of these bins contained more than 1\% of all pairs and
had $t_\eff\leq 2\,$days, we eliminated
the entire bin.  That is, we took the very high number of candidates
in a bin as an indication that they had a common cause in the observing
conditions, such as interplay between a bright moon and camera optics.
It is possible that a handful of real microlensing events were eliminated
in this way, but the method was essential to eliminating several hundred
thousand spurious candidates (given that we had not yet invented
the ``short-timescale artifact'' remover).

\subsection{{Periodic-Variable Finder}
\label{sec:periodic-var}}

The term ``periodic variable'' conveys two distinct ideas.  Normally,
one considers that the profile of its ``variation'' should repeat
(or approximately repeat) according to some regular (or nearly
regular) ``period''.  While there is no shortage of such variables,
there are also stars that vary quite regularly, but show very 
different amplitudes of variation (and sometimes forms of variation)
at each cycle.  We therefore adapted the event finder to identify
both types.

Recall that, as originally constructed, the event finder would
fit two flux parameters to each of several thousand different
magnification profiles that are defined by $(t_0,t_\eff)$, i.e., the
time of peak magnification and the effective timescale.  The
fits are restricted to data lying within $t_0\pm 5\,t_\eff$.  In the event
finder itself, 
only the best fit is kept.  The periodic-variable finder instead
keeps the top 50 $(t_0,t_\eff)$ pairs and then restricts consideration
to those with $\Delta\chi^2>0.2\Delta\chi^2_\max$.

It then attempts to find ``periodic behavior'' in a subset of the 50
cataloged $(t_0,t_\eff)$ pairs.  This subset is defined as the profiles for
which $t_\eff=t_{\eff,\max}$,
where $t_{\eff,\max}$ is the effective timescale
of the best-fit profile.  First, ``peaks'' (in $t_0$)
are successively identified within this subset.  The first ``peak'' 
is simply the $(t_0,t_{\eff,\max})$ with the highest $\Delta\chi^2$.
All pairs $(t_0,t_{\eff,\max})$ within $t_{\eff,\max}$ of this
peak are then grouped with it and eliminated from further consideration,
and the second ``peak'' is chosen as the $(t_0,t_{\eff,\max})$ with the 
highest $\Delta\chi^2$ among those remaining.  This process is
repeated until the $(t_0,t_{\eff,\max})$ subset is exhausted.

Then, the
algorithm checks to see if there are multiple peaks with the same period.
Based on a review of real periodic variables
(of all types), we restrict consideration to periods $P$ satisfying
$5< P/t_{\eff,\max}<13$.  That is, if the peak of a periodic variable
is fitted to a microlensing profile, the fit will typically have an effective
timescale that is about 12\% of the period, with some variation.  If
there are three or more cataloged $(t_0,t_{\eff,\max})$ pairs that are
matched within $t_{\eff,\max}$ of a predicted peak for a given trial period
$P$, then we exclude the star as a
``periodic variable''.  If there are only two $(t_0,t_{\eff,\max})$ pairs,
then we conduct two additional tests, both of which begin by identifying
the ``best'' period, i.e., the one that produces the closest match
to the second peak of $t_{\eff,\max}$ profiles.

First, we check whether there are other $t_\eff$ (i.e., other than
$t_{\eff,\max}$) for which there are three or more peaks that line up
with the ``best'' period derived from the $(t_0,t_{\eff,\max})$ profiles.
If so, we eliminate the star.  If not, then we ask whether
$(P,t_{\eff,\max})$ satisfy two further conditions.
The first condition
is that $P$ lies in the restricted range $6< P/t_{\eff,\max}<10$, which is
populated by a substantial majority of real variables.  Second,
we demand $P>T/4$, where $T$ is the duration of the season.  This
guarantees first 
that the interval between peaks is quite long, i.e., $\ga 60\,$days,
and second that no more than one peak from the periodic variable is
``missed'' (due to low amplitude).

Elimination of such two-peak ``periodic'' variables may seem dangerous
at first sight because any two peaks would seem to satisfy the criteria.
Recall, however, that the range of periods is quite restricted.   While
it is not difficult to construct a wide-binary microlensing event that
satisfies these criteria, the fraction of random events that actually do satisfy
them is quite small.  The second component of the binary must yield
a fit with the same $t_\eff$ as the first with $\geq 20\%$ of its $\Delta\chi^2$,
and the separation between peaks must be in a narrow range determined
by this $t_\eff$.  Moreover, the inferred period must be at least
$P\ga 60\,$days.
This mode of rejection is quite effective at rejecting long-period
variables.  It does induce a small amount of risk, and for this
reason it may be eliminated in some future year when the great
majority of such long-period variables are already cataloged.

\subsection{{Final Vetting}
\label{sec:final-vet}}

After all ``clear'' and ``possible'' microlensing events were selected,
we performed two additional inspections.  First, we had developed a special
pipeline to generate high-quality pySIS \citep{albrow09} reductions,
which was mainly for the purpose of the data release 
(see Section~\ref{sec:product}).  However, inspection of these higher-quality
reductions was also useful to identify variables of various kinds,
particularly CVs.

Second, we examined the combined 2016-2017 DIA light curves.  It was not
our original intention to do this because it obviously required waiting
until most or all of the 2017 season was over.  However, by the time
the pySIS pipeline was debugged and implemented, all 2017 DIA reductions
were complete.  In fact, these combined light curves proved quite
valuable in identifying long-period variables as well as repeating 
variables, probably mostly CVs.
Indeed, because quite a few of these ``repeaters'' had only one other
outburst within the two years (actually, two seasons) of data, it is very
likely that there are other ``repeaters'' masquerading as microlensing
events, whose frequency of repetition is yet lower.

In all, the event finder found 3,045,815 candidates, 
which the friends-of-friends
algorithm grouped into 1,123,819 groups.  102,121 of these were eliminated as
variables by matching to published catalogs or to 2015 KMTNet variables
or by the periodic-variable finder.  601,174 were eliminated by matching
to previous artifacts or by one of the two variants of artifact finder
discussed above.  This left 420,524 candidates to be viewed by operator.
Among these, 2065 were chosen as ``clear microlensing'' and 532 as
``possible microlensing''.  Of these 134 proved to be duplicates.  Then
based on inspection of their 2016 pySIS 
light curves and their 2016-2017 DIA light curves, a further 300
candidates were eliminated, leaving 1856 ``clear microlensing'' and 307
``possible microlensing'' events.  A map of these events is shown in
Figure~\ref{fig:2016events}.

\section{{Data Products}
\label{sec:product}}

All the data for this release are available at
http://kmtnet.kasi.re.kr/ulens/ from which one can access
http://kmtnet.kasi.re.kr/ulens/event/2016/ .

For almost all events, one can access DIA and pySIS data in two forms,
pictorial and ascii data files.  In a few cases the automated
pySIS reductions failed.  The page of each event contains a finding chart
and the best fit $(t_0,t_\eff,u_0)$ parameters from the original search
as well as cross-identification to OGLE and/or MOA listings.

\section{{Data Policy}
\label{sec:policy}}

Our data policy is basically the same as for 2015
(which we also anticipate will be the long term policy), and we urge the
reader interested in using the data presented here to read that
policy in \citet{eventfinder}.  Here we re-emphasize only the most
essential points and describe the small deviations from the 2015
policy.

The central point is that all data presented in Section~\ref{sec:product}
will become free for public use as soon as this paper and all papers
that use these data and that have {\it already} been submitted are accepted
for publication.  Prior to that time, papers can be prepared for
publication using these data, but they cannot be submitted to journals nor
posted on arXiv.

We welcome users of these data to collaborate with the KMTNet team, 
but we do not insist on it.  In
particular, while the pySIS pipeline reductions are generally of very
high quality, they can in some cases be improved using a tender-loving-care
(TLC) approach.  Such TLC light curves could be one form of collaboration.
However, if other workers would like to do their own re-reductions
without our cooperation, we will send postage-stamps of all epochs,
provided that we are given solid evidence of advanced preparation of
a publishable paper.  If requested, such postage stamps would
also include $V$-band data.

The 13 submitted papers based on these data are
\citet{kb160212}, 
\citet{ob160168,ob161045}, 
\citet{ob160596}, 
\citet{ob160613,ob161469}, 
\citet{ob160693,ob161190}, 
\citet{ob160733,ob161003}, 
\citet{ob161067}, 
\citet{ob161195}, 
and
\citet{ob161266}.

Finally, we note that a subset of the 2016 data release is available
without restriction, namely the events that broadly overlap the
{\it Kepler K2}C9 campaign.  As discussed by \citet{2016k2}, these
are available at http://kmtnet.kasi.re.kr/ulens/event/2016k2/ .
These same light curves are also available at
http://kmtnet.kasi.re.kr/ulens/event/2016/ in order to maintain
an integrated record of 2016 KMTNet events.  However, we continue
to maintain the {\it K2}C9 page so there is an unambiguous record
of which events have unrestricted access.

\section{{Future Plans}
\label{sec:plans}}

For 2015 data, we released DIA-pipeline light curves roughly 16 months
after the close of the 2015 season.  We also set two goals for improvement:
first, release high-quality pySIS light curves; second, reduce
the delay of release to  roughly 6 months after the close of 
each microlensing season, i.e., roughly April of the following year.

With this release, we have met the first goal but have fallen further 
behind on the second, i.e., to 18 months after the close of the 2016 season.
Nevertheless, on the basis of ongoing practical experience, we believe that
the original goal will be achieved for the 2018 release.  In particular,
the 2017 event selection was completed in early January 2018.  However,
we could not begin high-quality reductions until April 2018 due to
a backlog of work on 2016 data, at which point most computing power
was already engaged in the processing of 2018 data.  All such backlogs 
should be cleared by the time 2018 event selection is complete.

\begin{figure}
\plotone{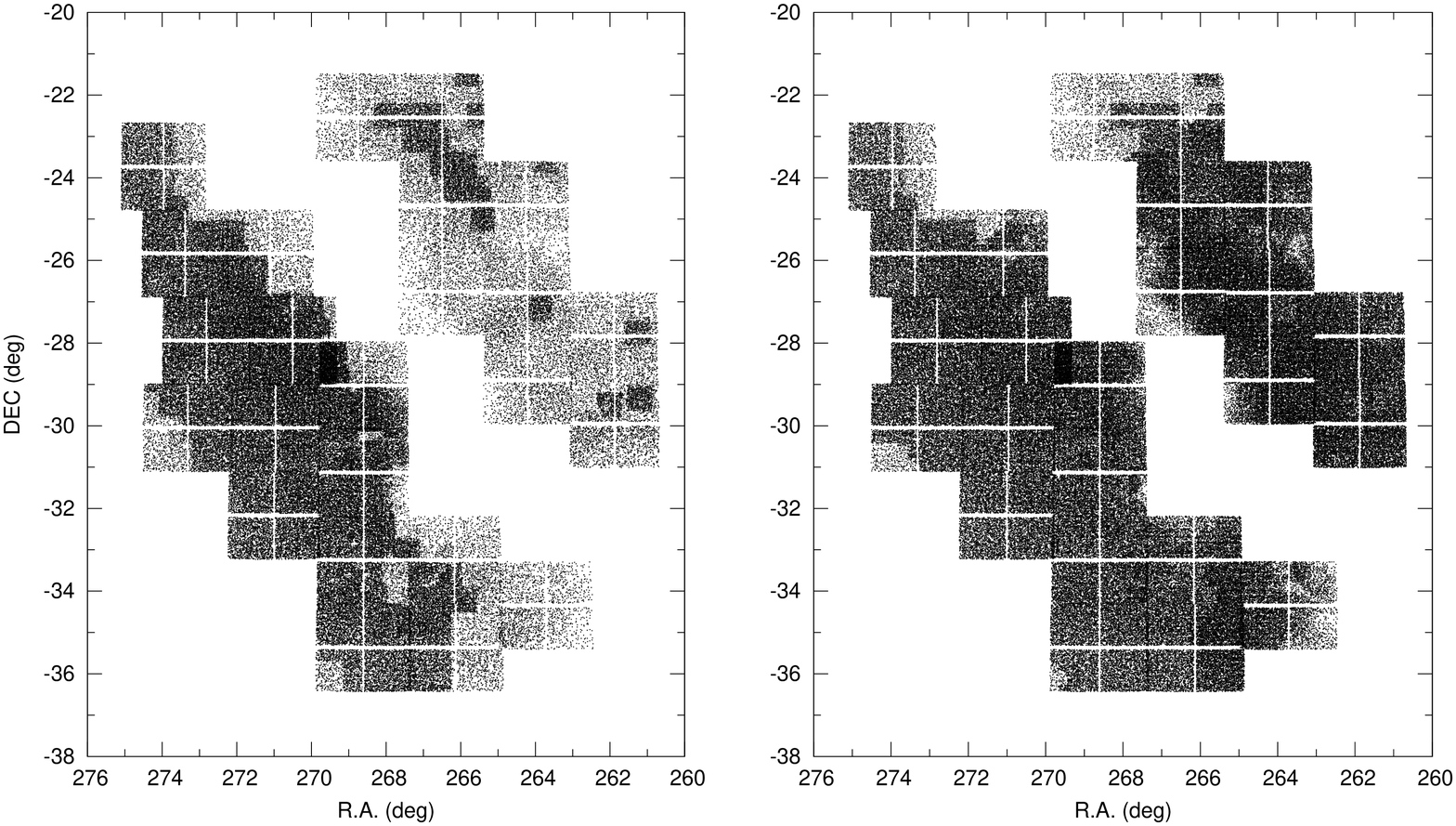}
\caption{Star catalog for 24 of the 27 KMTNet fields showing 1/2000 entries.
Fields  BLG41/42/43, which strongly overlap BLG01/02/03, are not shown
to avoid clutter, but these fill in the chip gaps of the latter three fields.
Left: Catalog actually used in 2016.  The regions covered by
the OGLE-III catalog \citep{oiiicat} are substantially denser than
those that are not and for which we constructed our own catalog
using DoPhot \citep{dophot}.  The boundaries between these are
sharp and rectilinear.  In addition, particularly near the Galactic
plane, there are regions of very low star-catalog density due to
high extinction.  See Figure~12 from \citet{eventfinder} for the
field-numbering scheme.
Right: Upgraded catalog that replaces the previous
DoPhot catalog with a CFHT catalog \citep{cfhtcat}, wherever the latter is
available.  This will be applied to 2018+ data going forward and
may be retrospectively applied to 2016 and 2017 data as well.
}
\label{fig:starcat}
\end{figure}

\begin{figure}
\plotone{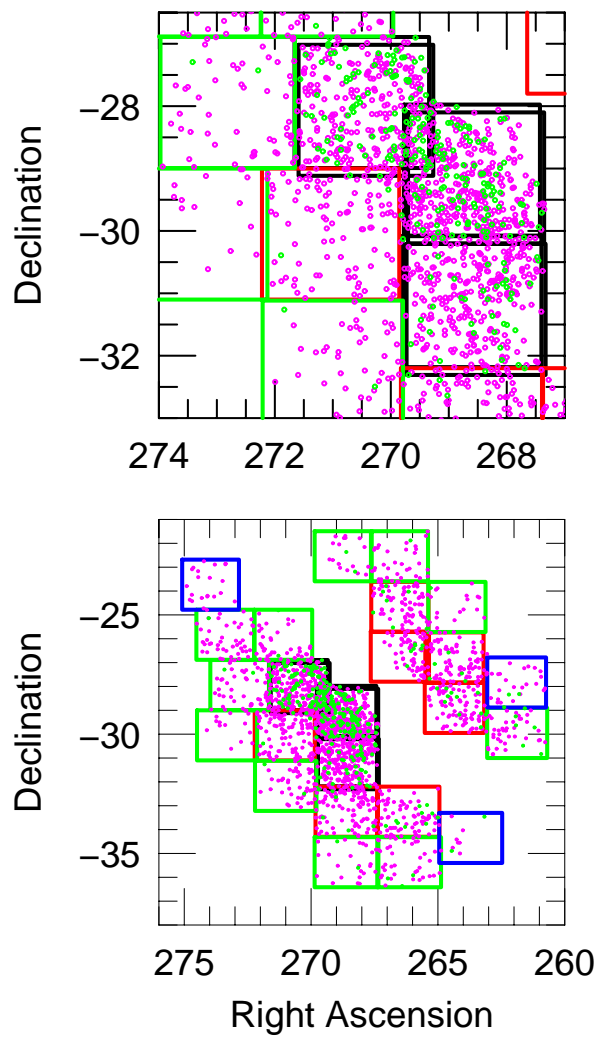}
\caption{Distribution in equatorial coordinates of the 2163
microlensing events found by KMTNet from its 2016 data, including
1856 ``clear'' (magenta) and 307 ``possible'' (green) events.  
The KMT fields are outlined and color-coded by cadence, with
(black, red, green, blue) corresponding to 
$\Gamma=(2,1,0.4,0.2)\,{\rm hr}^{-1}$.  See Figure~12 from 
\citet{eventfinder} for the field-numbering scheme.
The upper panel highlights the 6 overlapping high-cadence
fields (clockwise from upper left), BLG03/43, BLG02/42, BLG01/41 .
}
\label{fig:2016events}
\end{figure}

Finally, subsequent to the analysis reported here,
we have upgraded our star catalog by replacing our own 
DoPhot-based catalog with the CFHT-based 
catalog\footnote{http://decaps.skymaps.info} of \citet{cfhtcat} in
all regions that the two coincide.  This has increased the overall
size of the catalog from $3\times 10^8$ to $5\times 10^8$ stars.
See Figure~\ref{fig:starcat}.  The new catalog is already being
incorporated into the photometry of 2018 KMTNet data.  If resources
permit, we will reprocess the 2016 and 2017 data using this
new catalog as well.  In this case, we will update the 2016 webpage with
additional events.

\acknowledgments 
Work by YKJ and AG were supported by AST-1516842 from the US NSF.
IGS and AG were supported by JPL grant 1500811.
This research has made use of the KMTNet system operated by the Korea
Astronomy and Space Science Institute (KASI) and the data were obtained at
three host sites of CTIO in Chile, SAAO in South Africa, and SSO in
Australia.
Work by C.H. was supported by the grant (2017R1A4A101517) of
National Research Foundation of Korea.
Work by YS was supported by an appointment to the NASA Postdoctoral
Program at the Jet Propulsion Laboratory, California Institute of
Technology, administered by Universities Space Research Association
through a contract with NASA.

\end{document}